\documentclass[12pt]{article}

\usepackage{graphics,epsfig,here,color,cite,axodraw}

\definecolor{Blue}{named}{Blue}
\definecolor{Red}{named}{Red}
\definecolor{Green}{named}{ForestGreen}
\definecolor{Black}{named}{Black}
\definecolor{Olive}{named}{OliveGreen}
\definecolor{Royal}{named}{RoyalBlue}
\definecolor{Orange}{named}{YellowOrange}
\definecolor{Yellow}{named}{Goldenrod}
\definecolor{Cornblue}{named}{CornflowerBlue}
\definecolor{Lila}{named}{DarkOrchid}

\def\beq   {\begin{equation}}
\def\eeq   {\end{equation}}
\def\beqd  {\begin{displaymath}}
\def\eeqd  {\end{displaymath}}
\def\beqaa {\begin{eqnarray}}
\def\eeqaa {\end{eqnarray}}

\def\ti  {\tilde}

\def\sz{\ifmmode{\tilde{\chi}^0} \else{$\tilde{\chi}^0$} \fi}
\def\sw{\ifmmode{\tilde{\chi}} \else{$\tilde{\chi}$} \fi}

\newcommand{\lsim}{\;\raisebox{-0.9ex}{$\textstyle\stackrel{\textstyle<}
           {\sim}$}\;}

\newcommand{\be}[1]{\begin{equation} \label{(#1)}}
\newcommand{\ee}{\end{equation}}
\newcommand{\baq}[1]{\begin{eqnarray} \label{(#1)}}
\newcommand{\eaq}{\end{eqnarray}}
\newcommand{\rf}[1]{(\ref{(#1)})}
\newcommand{\ba}{\begin{array}}
\newcommand{\ea}{\end{array}}

\newcommand{\slashed}[1]{\not\!#1}

\begin{document}

\pagestyle{empty}

\vspace*{-1cm} 
\begin{flushright}
hep-ph/0610431 \\
UWThPh-2006-27 \\
IPPP/06/73\\
DCPT/06/146
\end{flushright}

\vspace*{1cm}

\begin{center}

{\Large
{\bf Selectron production at an $e^-e^-$ linear collider
with transversely polarized beams
}}

\vspace{2cm}

{\large 
A.~Bartl$^{1}$, H.~Fraas$^{2}$, K.~Hohenwarter-Sodek$^{1}$, 
T.~Kernreiter$^{1}$, G.~Moortgat-Pick$^{3}$, A.~Wagner$^{2}$}
\end{center}
\vspace{1cm}

{\it \noindent $^{1}$~Institut f\"ur Theoretische Physik, Universit\"at Wien, A-1090
Vienna, Austria\\
$^{2}$~Institut f\"ur Theoretische Physik und Astrophysik, Universit\"at W\"urzburg,\\ 
\phantom{$^{2}$ }D-97074 W\"urzburg, Germany\\
$^{3}$~ IPPP, University of Durham, Durham DH1 3LE, UK}

\vspace{1.5cm}

\begin{abstract}
 We study selectron production at an $e^-e^-$ linear collider. 
 With the help of transverse beam polarizations, we define 
 CP sensitive observables in the production process
 $e^-e^-\to\ti e^-_L \ti e^-_R$. This process proceeds via 
 $t-$channel and $u-$channel exchange of neutralinos, and is 
 sensitive to CP violation in the neutralino sector.
 We present numerical results and estimate the significances 
 to which the CP sensitive observables can be measured.
 \end{abstract}

\newpage
\pagestyle{plain}

\section{Introduction}

Supersymmetry (SUSY) is one of the most attractive 
extensions of the Standard Model (SM).  
If SUSY particles are found at Tevatron or LHC, 
then one of the most important goals of 
the future international linear collider (ILC) will 
be the precise determination of the 
quantum numbers, masses and couplings of supersymmetric particles 
\cite{Aguilar-Saavedra:2005pw}.
In addition to the $e^+e^-$ mode of the ILC also the $e^- e^-$ mode
offers the possibility to study properties of
selectrons and neutralinos \cite{Feng:1999zv}. 

In this paper we investigate the potential of $e^- e^-$ collisions
with transverse beam polarizations for the determination of SUSY CP phases.
Our framework is the minimal supersymmetric standard model (MSSM) with 
complex parameters.
The usefulness of transverse beam polarizations at the ILC has been discussed
before for various observables \cite{TP,TPsusy,TPchar,TPNeu1,TPNeu2}.

We study the production process
\be{eq:slprod}
e^-e^-\to\ti e^-_L \ti e^-_R
\ee
which proceeds via neutralino exchange in the 
$t-$channel and $u-$channel 
\footnote{The amplitude squared of
selectron pair production $\tilde{e}^-_L\tilde{e}^-_L$, $\tilde{e}^-_R\tilde{e}^-_R$ 
does not depend on the transverse beam polarizations (see section \ref{crosssection}).}. 
Therefore, it is sensitive to the complex parameters in the
neutralino sector. These are (after reparametrization of the fields) the 
higgsino mass parameter $\mu$ and the $U(1)$ gaugino mass parameter $M_1$.
The current experimental bounds on the electric dipole
moments of electron, neutron and the atoms $^{199}$Hg and $^{205}$Tl 
suggest that the phase of $\mu$, $\phi_\mu$, may be more
restricted than the phase of $M_1$, $\phi_{M_1}$ 
(see for instance \cite{Barger:2001nu}). 
These constraints, however, are rather model
dependent \cite{Bartl:2003ju}. Therefore, it is necessary to determine 
the phases of the complex SUSY parameters by measurements of suitable 
CP sensitive observables.

We propose T-odd observables in the production process \rf{eq:slprod} 
by means of the azimuthal angular distribution of the selectrons. 
These observables require both electron beams to be transversely polarized.
Without transverse beam polarization no T-odd terms
involving a triple product correlation appear in 
the matrix element squared of the 
production process \rf{eq:slprod} due to the lack of 
three linearly independent momentum and/or polarization vectors.
This remains true if the subsequent decays of the selectrons 
are taken into account, because the selectrons are scalar particles.
We stress that in the reaction $e^+e^-\to\ti e^+_i \ti e^-_j$ 
even for transversely polarized $e^+$ and $e^-$ beams no useful
CP sensitive observable can be found since in this case the CP sensitive terms
are proportinal to the tiny left-right selectron mixing.

The complex parameters $M_1$ and/or $\mu$
(with $\phi_{M_1}$ and/or $\phi_\mu\neq 0,\pi$)
give rise to the T-odd observables to be considered.
A measurement of these T-odd observables therefore
allows us to obtain information on the
MSSM parameters, in addition to those which can be obtained
from a measurement of suitable T-odd observables in the production process
$e^+e^-\to\ti\chi^0_i\ti\chi^0_j$, $i,j=1,\dots,4$ 
\cite{Bartl:2004jj,TPNeu1,TPNeu2,Barger:2001nu,Choi:1999cc,Drees}.
Due to CPT invariance, at tree-level, these T-odd observables are actually 
CP sensitive observables.

In Section \ref{crosssection} we outline the calculation
of the production cross section for $e^-e^-\to\ti e^-_L \ti e^-_R$
with arbitrary beam polarizations.
In Section \ref{observables} we define our CP sensitive observables.
We present numerical results in Section \ref{numerics}, where we 
also estimate the measurability of the CP sensitive observables.
We summarize in Section \ref{conclusion}.

\section{Cross section \label{crosssection}}

\begin{figure}[t]
\begin{picture}(180,95)(-25,-60)
\SetWidth{1.5} 
\ArrowLine(0,60)(65,45)
\Line(63.5,45)(63.5,0)
\Vertex(65,45){2}
\Line(66.5,45)(66.5,0)
\Vertex(65,0){2}
\ArrowLine(0,-15)(65,0)
\DashArrowLine(65,45)(130,60){3}
\DashArrowLine(65,0)(130,-15){3}
\Text(8,70)[]{{\large $e^-(p_1,s_1)$}}
\Text(8,3)[]{{\large $e^-(p_2,s_2)$}}
\Text(78,24)[]{{\large $\tilde\chi^0_k$}}
\Text(125,70)[]{{\large $\tilde e^-_i(p_{\tilde e_i})$}}
\Text(125,0)[]{{\large $\tilde e^-_j(p_{\tilde e_j})$}}

\end{picture}

\begin{picture}(180,95)(-250,-155)
\SetWidth{1.5}
\ArrowLine(0,60)(65,45)
\Line(63.5,45)(63.5,0)
\Vertex(65,45){2}
\Line(66.5,45)(66.5,0)
\Vertex(65,0){2}
\ArrowLine(0,-15)(65,0)
\DashArrowLine(65,45)(130,-10){3}
\DashArrowLine(65,0)(130,60){3}
\Text(8,70)[]{{\large $e^-(p_1,s_1)$}}
\Text(8,3)[]{{\large $e^-(p_2,s_2)$}}
\Text(52,24)[]{{\large $\tilde\chi^0_k$}}
\Text(145,70)[]{{\large $\tilde e^-_i(p_{\tilde e_i})$}}
\Text(145,0)[]{{\large $\tilde e^-_j(p_{\tilde e_j})$}}
\end{picture}
\vskip-4.5cm
\caption{Feyman graphs for selectron production in $e^-e^-$-collisions. 
\label{fig:prodfeynman}}
\end{figure}
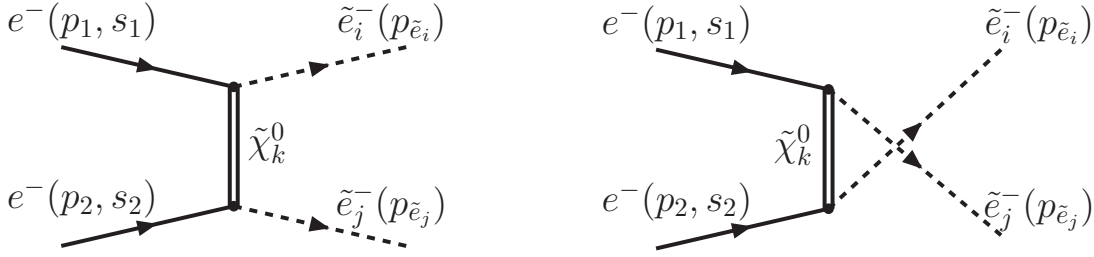

In the following we outline the calculation of the production cross section 
for $e^-e^-\to\ti e^-_i \ti e^-_j$, $i,j=1,2$, 
for arbitrary beam polarizations, 
neglecting the mass of the electron. $\ti e^-_i$, $i=1,2$, is the selectron
mass eigenstate, $m_{\ti e_1}<m_{\ti e_2}$.
The selectron mixing angle is 
$\theta_{\ti e}= 0~(\pi/2)$ for $\ti e_1 = \ti e_L$ ($\ti e_R$).
We first calculate the amplitude squared 
for process \rf{eq:slprod}.
The relevant part of the Lagrangian is given by
\be{eq:Lagrange}
{\mathcal L}_{\ti\chi\ti e e}=g~\bar{e}~(a_{kl}P_R+b_{kl}P_L)~\ti\chi^0_k~ 
\ti e^-_l+{\rm h.c.}~,
\ee
with $P_{L,R}=1/2(1\mp \gamma_5)$ and $g$ being the 
$SU(2)$ weak coupling constant.
The couplings $a_{kl}$ and $b_{kl}$ in Eq.~\rf{eq:Lagrange}
contain the neutralino mixing elements $N_{kl}$ and are given
in the basis ($\ti B,\ti W^3,\ti H_1^0,\ti H_2^0$) \cite{Haber:1984rc} as
\be{eq:couplings}
a_{k1}= \cos\theta_{\ti e} f^L_k~,\quad 
a_{k2}= -\sin\theta_{\ti e} f^L_k~,\quad 
b_{k1}= \sin\theta_{\ti e} f^R_k~,\quad 
b_{k2}= \cos\theta_{\ti e} f^R_k~,
\ee
with
\be{eq:couplings2}
f^L_k=\frac{1}{\sqrt{2}}(N_{k2}+\tan\theta_W N_{k1})~,\qquad 
f^R_k=\sqrt{2}\tan\theta_W N^*_{k1}~,
\ee
where $\theta_W$ denotes the weak mixing angle.
The amplitudes for $e^-e^-\to\ti e^-_i \ti e^-_j$ are
\be{eq:amplitude}
M_{ij}=M^t_{ij}+M^u_{ij}~,
\ee
see Fig.~\ref{fig:prodfeynman},
where $M^t_{ij}$ is the contribution from neutralino exchange
in the $t-$channel,
\be{eq:tchannel}
M^t_{ij}=g^2 \sum^4_{k=1} \Delta^t_k \bar{v}(p_2,s_2)(a_{ki}^* P_L +b_{ki}^*P_R)
(\slashed{p_1}-\slashed{p_{\ti e_j}}+m_k)(a_{kj}^* P_L +b_{kj}^*P_R)u(p_1,s_1)~,
\ee
and $M^u_{ij}$ is the $u-$channel neutralino exchange contribution,
\be{eq:uchannel}
M^u_{ij}=g^2 \sum^4_{k=1} \Delta^u_k \bar{v}(p_2,s_2)(a_{kj}^* P_L +b_{kj}^*P_R)
(\slashed{p_1}-\slashed{p_{\ti e_i}}+m_k)(a_{ki}^* P_L +b_{ki}^*P_R)u(p_1,s_1)~,
\ee
where $\Delta^t_k=i/((p_1-p_{\ti e_j})^2-m_k^2)$, 
$\Delta^u_k=i/((p_1-p_{\ti e_i})^2-m_k^2)$, $m_k$ denotes the neutralino masses,
$p_1$ and $p_2$ are the 4-momenta of the incoming electrons and
$p_{\ti e_i}$ is the 4-momentum of the corresponding selectron.

In the treatment of beam polarizations we use the covariant
projection operators \cite{TPchar,Renard} (for a different treatment see 
for instance \cite{TPNeu1}). In the limit of vanishing electron masses 
they read
\be{eq:proje1}
\sum_{s_1} {\bar u}(p_1,s_1)u(p_1,s_1)=
\frac{1}{2}({\bf 1}+P^1_L\gamma_5+\gamma_5 P^1_T \slashed t_1)\slashed p_1
\ee
and
\be{eq:proje2}
\sum_{s_2} {\bar v}(p_2,s_2)v(p_2,s_2)=
\frac{1}{2}({\bf 1}-P^2_L\gamma_5+\gamma_5 P^2_T
\slashed t_2)\slashed p_2~,
\ee
where $t_{1,2}$ are the transverse beam
polarization 4-vectors of the $e^-$ beams.  
In Eqs.~\rf{eq:proje1}  and \rf{eq:proje2}
$P^{1,2}_L$ [$-1\leq P^{1,2}_L \leq 1$] denote the degree
of the longitudinal polarizations of the $e^-$ beams and 
$P^{1,2}_T$ [$0\leq P^{1,2}_T \leq 1$] 
denote the degree of transverse polarizations, 
statisfying $(P^{1,2}_L)^2+(P^{1,2}_T)^2\leq 1$.

The amplitude squared for the production process 
$e^-e^-\to\ti e^-_i \ti e^-_j$ can be written as
\be{eq:ampsquared}
|M_{ij}|^2=|M^t_{ij}|^2+|M^u_{ij}|^2+2 \Re e\{M^t_{ij}M^{u\dagger}_{ij}\}~,
\ee
where in the following we only give the result for the production 
of different mass eigenstates 
\footnote{When neglecting selectron mixing, the result for $i = j$ 
is the same as given in \cite{Blochinger:2002zw} for 
longitudinal beam polarizations, 
i.e. the amplitude squared does not depend on the transverse beam polarizations
in this case.}
, i.e. $i\neq j$, because otherwise only the absolute values of the couplings enter 
and no CP-odd term appears in the amplitude squared.  
As was shown in \cite{Thomas:1997ng} the cross section for 
$\tilde e^-_L \tilde e^-_L$ production, although it is a CP-even observable, 
is quite sensitive to CP violation in the neutralino sector,
representing a complementary observable.   

We introduce a coordinate system by choosing 
the $z$-axis along the $\vec{p}_1$ direction in the c.m. system,
and $x$ and $y$ corresponding to a right-handed coordinate system.
In this coordinate system the transverse beam polarization 4-vectors 
in Eqs.~\rf{eq:proje1} and \rf{eq:proje2} are 
\be{eq:transvec}
t^{1,2}=(0,\cos\phi_{1,2},\sin\phi_{1,2},0)~.
\ee
We first consider the case $e^-e^-\to\ti e^-_1\ti e^-_2$, 
where $\ti e_1=\ti e_R$ and $\ti e_2=\ti e_L$.
We obtain
\be{eq:Msqtt}
|M^t_{12}|^2=\frac{g^4}{4}~
s~q^2~\sin^2\theta~c_{+-} 
\sum^4_{k,l=1} f^{L*}_k f^{L}_l f^{R*}_k f^{R}_l 
\Delta^{t}_k \Delta^{t*}_l~, 
\ee
\be{eq:Msquu}
|M^{u}_{12}|^2=\frac{g^4}{4}~
s~q^2~\sin^2\theta~c_{-+} 
\sum^4_{k,l=1} f^{L*}_k f^{L}_l f^{R*}_k f^{R}_l 
\Delta^{u}_k \Delta^{u*}_l~, 
\ee
\baq{eq:Msqut}
2 \Re e\{M^t_{12}M^{u\dagger}_{12}\}
&=&\frac{g^4}{2}~P^1_T~P^2_T~s~ q^2 \sin^2\theta
\sum^4_{k,l=1} \Delta^t_k \Delta^{u*}_l \nonumber \\ 
&&{}\times
\Re e\{f^{L*}_k f^{L}_l f^{R*}_k f^{R}_l [\cos(\eta-2\phi) 
- i \sin(\eta-2\phi)] \}~,
\eaq
where $c_{\pm\mp}=(1\pm P^1_L)(1\mp P^2_L)$, 
$q=\lambda^{1/2}(s,m^2_{\ti e_1},m^2_{\ti e_2})/(2 \sqrt{s})$, 
$E_{\ti e_{1,2}}=(s+m^2_{\ti e_{1,2}}-m^2_{\ti e_{2,1}})/(2 \sqrt{s})$, 
$m_{\ti e_i}$ are the selectron masses, $\eta=\phi_1+\phi_2$, 
$\theta$ and $\phi$ being the polar angle and azimuthal angle
of $\ti e^-_2$.
For the case $\ti e_1=\ti e_L$, $\ti e_2=\ti e_R$, the amplitude squared is obtained
by the replacements $c_{+-}\to c_{-+}$ in Eq.~\rf{eq:Msqtt} and
$c_{-+}\to c_{+-}$ in Eq.~\rf{eq:Msquu}, 
and by changing the overall sign in Eq.~\rf{eq:Msqut}.

The differential cross section for $e^-e^-\to\ti e^-_1 \ti e^-_2$ is given by
\be{eq:crosssection}
\frac{d\sigma}{d\Omega}=\frac{1}{8(2\pi)^2}\frac{q}{s^{3/2}}|M_{12}|^2~,
\ee
with $d\Omega=\sin\theta d\theta d\phi$ and $|M_{12}|^2$ as given 
in Eq.~\rf{eq:ampsquared}.
Note that the production cross section $\sigma$ 
is independent of the transverse beam 
polarizations, because the appropriate contributions in the amplitude squared
depend on $\cos(\eta-2\phi)$ or on $\sin(\eta-2\phi)$, 
see Eq.~\rf{eq:Msqut}, and vanish 
if integrated over the whole range of the azimuthal angle $\phi$.

\section{CP sensitive observables\label{observables}}

In this section we define our CP sensitive observables for the
production process $e^-e^-\to\ti e^-_L\ti e^-_R$ with transverse
$e^-$ beam polarizations.
By inspecting Eq.~\rf{eq:Msqut}, we observe that the CP sensitive term 
which involves the imginary part of the couplings 
$f^{L*}_k f^{L}_l f^{R*}_k f^{R}_l$
is proportional to 
\be{eq:propadiff}
\sum^4_{k<l}(\Delta^{t}_k \Delta^{u}_l-\Delta^{t}_l \Delta^{u}_k)~
\Im m\{f^{L*}_k f^{L}_l f^{R*}_k f^{R}_l\}~,
\ee
and would be zero when integrated over the whole range of the
polar angle $\theta$ because of the symmetry of the propagator term.
We therefore have to divide the integration over 
$\theta$ into two regions \cite{TPNeu1}. This amounts 
to a sign change of $\cos\theta$,
which we can take into account by multiplying \rf{eq:propadiff} by a 
weight function
\be{eq:f}
{\mathcal H}_1={\rm sign}[\sin(\eta-2\phi)\cos\theta]~.
\ee
An other choice of the weight function can be given 
by matching the angular dependence of the term of interest 
in the amplitude squared \cite{TPNeu2,Atwood:1991ka}.
This can be achieved with the weight function
\be{eq:fopt}
{\mathcal H}_2=\sin(\eta-2\phi)\cos\theta\sin^2\theta~.
\ee
As our CP sensitive observables, we define the expectation values
of ${\mathcal H}_i, i=1,2$, given as
\be{eq:Obs}
\langle {\mathcal H}_i\rangle=\frac{1}{\sigma}\int d\Omega 
\frac{d\sigma}{d\Omega}~{\mathcal H}_i~.
\ee
Due to the requirement that the statistical error of the observable 
should not exceed its size, we have
\be{eq:Error}
\frac{|\langle {\mathcal H}_i\rangle|}{\Delta\langle {\mathcal H}_i\rangle}>1~,
\ee
where $\Delta\langle {\mathcal H}_i\rangle={\mathcal N}_\sigma/\sqrt{N}
\sqrt{\langle {\mathcal H}_i^2\rangle-\langle {\mathcal H}_i\rangle^2}\simeq 
{\mathcal N}_\sigma/\sqrt{N}\sqrt{\langle {\mathcal H}_i^2\rangle}$, 
with ${\mathcal N}_\sigma$ being the 
number of standard deviations and $N=\sigma {\mathcal L}$ the number of events,
where ${\mathcal L}$ denotes the integrated luminosity.
Using Eq.~\rf{eq:Error}, we define an effective CP observable 
given as
\be{eq:EffAsy}
{\hat O}[{\mathcal H}_i]=\sqrt{\sigma}~
\frac{\langle {\mathcal H}_i\rangle}{\sqrt{\langle {\mathcal H}_i^2\rangle}}~.
\ee
$|{\hat O}[{\mathcal H}_i]|\cdot \sqrt{{\mathcal L}}$ is then the number of 
standard deviations to which the corresponding observable, Eq.~\rf{eq:Obs}, 
can be determined to be non-zero.
 
Note that a measurement of the CP sensitive observables
disussed above requires the reconstruction of the 
production plane. If all masses involved are known, this can be accomplished 
either in a unique way or with a two-fold ambiguity, depending on 
the decay pattern of the produced selectrons \cite{TPNeu1,TPNeu2}. 

\section{Numerical results \label{numerics}}

\begin{figure}
\setlength{\unitlength}{1mm}
\begin{center}
\begin{picture}(150,35)
\put(-53,-152.5){\mbox{\epsfig{figure=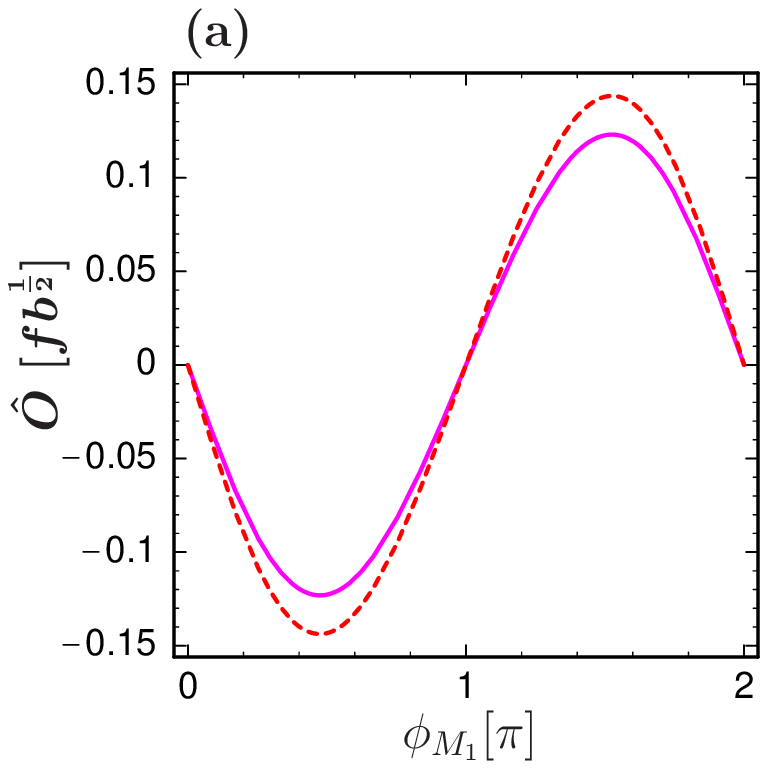,height=22.cm,width=19.4cm}}}
\put(32,-137){\mbox{\epsfig{figure=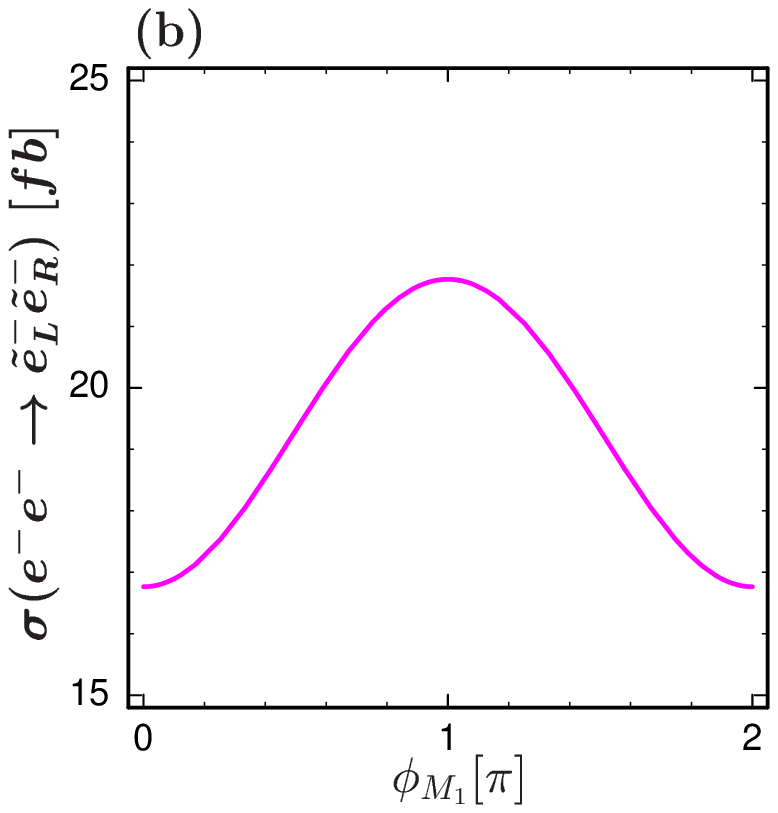,height=20.1cm,width=17.7cm}}}
\end{picture}
\end{center}
\caption{(a) Effective CP observable $\hat{O}[{\mathcal H}_i]$, Eq.~\rf{eq:EffAsy},
as a function of $\phi_{M_1}$ for scenario B
of Table~\ref{tab1}, 
with ${\mathcal H}_1=\rm{sign}[\cos\theta \sin(\eta-2\phi)]$ (solid line) and
${\mathcal H}_2=\sin^2\theta \cos\theta\ \sin(\eta-2\phi)$ (dashed line)
and (b) the 
corresponding cross section $\sigma(e^-e^-\to\ti e^-_L\ti e^-_R)$. }
\begin{table}[H]
\hspace{0.5cm}   \begin{tabular}{|l|r|r|r|r|}
		 \hline
			\textbf{A}
		  & $\tilde\chi^0_1$ & $\tilde\chi^0_2$ & $\tilde\chi^0_3$ & 
                    $\tilde\chi^0_4$ \\
		 \hline\hline
		  $\tilde B$    & 0.036 & 0.010 & 0.953 & 0.000 \\
		  $\tilde W^3$  & 0.023 & 0.007 & 0.003 & 0.967 \\
		  $\tilde H_1$  & 0.515 & 0.473 & 0.007 & 0.005 \\
		  $\tilde H_2$  & 0.426 & 0.510 & 0.036 & 0.028 \\
		 \hline\hline
			 Mass & 100.8 & 118.0  & 259.2  & 514.4 \\
		 \hline
		 \end{tabular}
		\begin{tabular}{|l|r|r|r|r|}
		\hline
			\textbf{B}
		 & $\tilde\chi^0_1$ & $\tilde\chi^0_2$ & $\tilde\chi^0_3$ & 
                   $\tilde\chi^0_4$ \\
		\hline\hline
			$\tilde B$         &  0.009 & 0.002  & 0.140 & 0.849 \\
			$\tilde W^3$       &  0.052 & 0.006  & 0.823 & 0.120 \\
			$\tilde H_1$       &  0.508 & 0.476  & 0.009 & 0.006 \\
			$\tilde H_2$       &  0.431 & 0.516  & 0.028 & 0.025 \\
		\hline\hline
			 Mass &  102.2 & 122.4 &  416.6 & 437.5 \\
		\hline
		\end{tabular}

\vspace{0.1cm}
\hspace{0.5cm}  \begin{tabular}{|l|r|r|r|r|}
		\hline
			\textbf{C}
		 & $\tilde\chi^0_1$ & $\tilde\chi^0_2$ & $\tilde\chi^0_3$ & 
                   $\tilde\chi^0_4$ \\
		\hline\hline
		 $\tilde B$    & 0.015 & 0.011 & 0.088 & 0.887 \\
		 $\tilde W^3$  & 0.336 & 0.013 & 0.619 & 0.033 \\
		 $\tilde H_1$  & 0.390 & 0.472 & 0.114 & 0.024 \\
		 $\tilde H_2$  & 0.259 & 0.505 & 0.180 & 0.056 \\
		\hline\hline
		   Mass & 106.2 & 162.7 & 251.4 & 311.2  \\
		\hline
		\end{tabular}
	\caption{\label{tab2} Neutralino compositions and mass 
                              spectra [GeV] for the scenarios A, B and C.}
\end{table}
\label{fig:plot2}
\vspace*{1.4cm}
\end{figure}

Now we analyze numerically the effective CP observables 
defined in Eq.~\rf{eq:EffAsy} for the reaction
$e^-e^-\to\ti e^-_L\ti e^-_R$ at a linear collider with $\sqrt{s}=500$~GeV
and transverse beam polarizations. We assume that a degree of transverse 
polarization of $90\%$ is feasible for each of the two electron beams.
Furthermore, in order to estimate the significance of a measurement
of the CP sensitive observables we assume that one third of the integrated 
luminosity ${\mathcal L}$ of the $e^+e^-$ mode can be achieved 
\cite{Feng:1999zv}. For our numerical analysis we choose three
scenarios, A, B and C, defined in Table~\ref{tab1}. In Table~\ref{tab2}
we give the masses and the compositions of the neutralinos 
$\ti{\chi}^0_i$ in these scenarios.
 
\begin{table}[H]
\begin{center}
\begin{tabular}{|c||c|c|c|c|c|c|c|c|} \hline
 Scenario 
& \multicolumn{1}{c|}{$|M_1|$} 
& \multicolumn{1}{c|}{$\phi_{M_1}$} 
& \multicolumn{1}{c|}{$M_2$} 
& \multicolumn{1}{c|}{$|\,\mu\,|$}
& \multicolumn{1}{c|}{$\phi_{\mu}$}  
& \multicolumn{1}{c|}{$\tan{\beta}$}  
& \multicolumn{1}{c|}{$m_{\tilde{e}_L}$}
& \multicolumn{1}{c|}{$m_{\tilde{e}_R}$}\\\hline\hline
 \textbf{A} 
& \multicolumn{1}{c}{250.5} 
& \multicolumn{1}{c}{$0.5\pi$} 
& \multicolumn{1}{c}{500} 
& \multicolumn{1}{c}{115}  
& \multicolumn{1}{c}{0}
& \multicolumn{1}{c}{5} 
& \multicolumn{1}{c}{200}
& \multicolumn{1}{c|}{170}\\\hline
 \textbf{B} 
& \multicolumn{1}{c}{430} 
& \multicolumn{1}{c}{$0.5\pi$} 
& \multicolumn{1}{c}{400} 
& \multicolumn{1}{c}{120}  
& \multicolumn{1}{c}{0}
& \multicolumn{1}{c}{3} 
& \multicolumn{1}{c}{160}
& \multicolumn{1}{c|}{130}\\\hline 
 \textbf{C} 
& \multicolumn{1}{c}{300} 
& \multicolumn{1}{c}{$0.5\pi$} 
& \multicolumn{1}{c}{200} 
& \multicolumn{1}{c}{160}  
& \multicolumn{1}{c}{0}
& \multicolumn{1}{c}{3} 
& \multicolumn{1}{c}{170}
& \multicolumn{1}{c|}{120}\\\hline 
\end{tabular}\\[0.5ex]
\vskip0.4cm
\caption{\label{tab1}
Input parameters $|M_1|$, $\phi_{M_1}$, $M_2$, $|\mu|$, $\phi_\mu$,
$m_{\tilde{e}_L}$ and $m_{\tilde{e}_R}$.
All mass parameters are given in GeV.}
\end{center}
\end{table}

\begin{figure}[t]
\setlength{\unitlength}{1mm}
\begin{center}
\begin{picture}(150,35)
\put(-53,-152.5){\mbox{\epsfig{figure=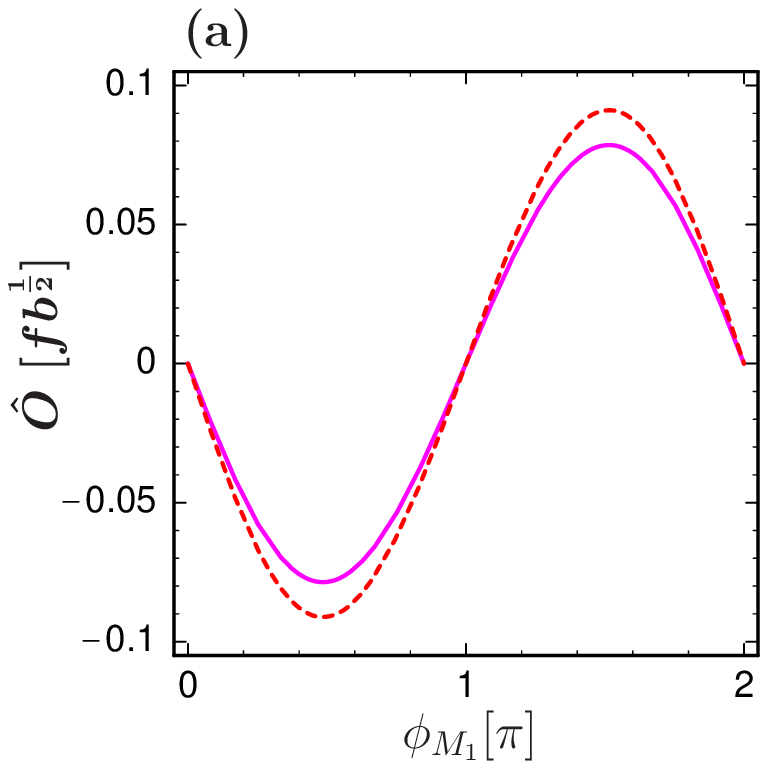,height=22.cm,width=19.4cm}}}
\put(32,-137){\mbox{\epsfig{figure=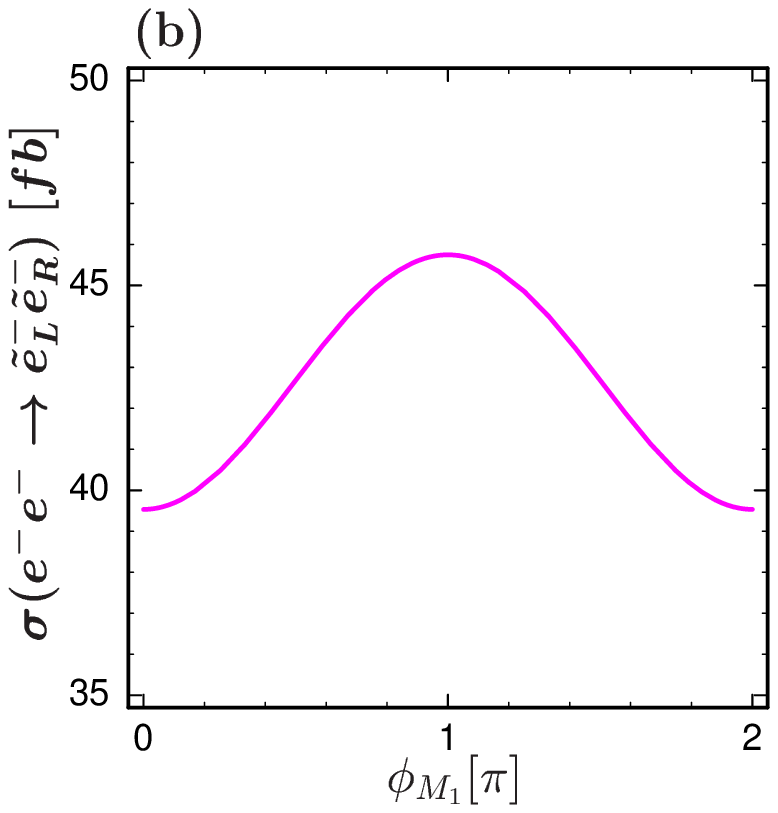,height=20.1cm,width=17.7cm}}}
\end{picture}
\end{center}
\caption{(a) Effective CP observable $\hat{O}[{\mathcal H}_i]$, Eq.~\rf{eq:EffAsy},
as a function of $\phi_{M_1}$ for scenario A 
of Table~\ref{tab1}, 
with ${\mathcal H}_1=\rm{sign}[\cos\theta \sin(\eta-2\phi)]$ (solid line) and
${\mathcal H}_2=\sin^2\theta \cos\theta\ \sin(\eta-2\phi)$ (dashed line)
and (b) the corresponding cross section 
$\sigma(e^-e^-\to\ti e^-_L\ti e^-_R)$. }
\label{fig:plot1}
\end{figure}

\noindent {\bf a) Case with $M_1/M_2$ GUT-relation}\\[.5em]
In Fig.~\ref{fig:plot1} we show the effective CP observables, 
Eq.~\rf{eq:EffAsy}, that are based on the weight functions ${\mathcal H}_1$ 
and ${\mathcal H}_2$ and the associated production cross section
as a function of $\phi_{M_1}$ for scenario A, given in Table~\ref{tab1}.
In this scenario we assume the GUT-inspired relation 
$|M_1|=5/3 \tan^2 \Theta_W M_2$.
Table~\ref{tab2} shows that in scenario A, 
$\ti{\chi}^0_1$ and $\ti{\chi}^0_2$ are mainly higgsinos,
$\ti{\chi}^0_3$ is mainly a bino and $\ti{\chi}^0_4$ is mainly a wino.
For the parameters chosen, the leading contribution to  
$\sigma(e^-e^- \to \ti{e}^-_L\ti{e}^-_R)$ 
stems from $\ti{\chi}^0_3$ exchange in the $t-$channel and $u-$channel,
since $\ti{e}_L$ couples to the bino and wino components
of the neutralinos and $\ti{e}_R$ to their bino component, 
see Eq.~\rf{eq:couplings2}.
On the other hand, because $\ti{\chi}^0_1$ has an 
appreciable bino component (3.6\%), the 
leading CP violating contribution to the CP sensitive observables, 
Eq.~\rf{eq:Obs}, is due to the interference
of the $\ti{\chi}^0_1$ and $\ti{\chi}^0_3$ 
exchange amplitudes, Eq.~\rf{eq:Msqut}. 
In Figs.~\ref{fig:plot1}a and b we can clearly see the antisymmetric
dependence of the CP sensitive observables on the phase $\phi_{M_1}$ 
while the production cross section is symmetric in $\phi_{M_1}$. 
Thus, both kinds of observables 
are needed for an unambiguous determination of $\phi_{M_1}$.
However, in order to probe the CP sensitive observables $\langle {\mathcal H}_1\rangle$ and 
$\langle {\mathcal H}_2\rangle$ at $3\sigma$, integrated luminosities of
$\mathcal{L}=1500$~fb$^{-1}$ and $\mathcal{L}=1042$~fb$^{-1}$ would be required for
scenario A.

\noindent {\bf b) Case without $M_1/M_2$ GUT-relation}\\[.5em]
In Fig.~\ref{fig:plot2} we plot the effective CP observables, 
Eq.~\rf{eq:EffAsy}, that are based on the weight functions ${\mathcal H}_1$ 
and ${\mathcal H}_2$ and the associated production cross section
as a function of $\phi_{M_1}$ for scenario B, given in Table~\ref{tab1}.
In scenario B, $\ti{\chi}^0_1$ and $\ti{\chi}^0_2$ are
again mainly higgsinos (see Table~\ref{tab2}), 
however, we do not assume the GUT-relation 
between the gaugino mass parameters $|M_1|$ and $M_2$. 
For scenario B the maximum (minimum) values of the effective CP observables
are reached at $\phi_{M_1}\approx 0.5\pi~(1.5\pi)$.
The integrated luminosity required for a measurement 
of the associated CP sensitive 
observables $\langle {\mathcal H}_1\rangle$ and 
$\langle {\mathcal H}_2\rangle$ at $3\sigma$ is $\mathcal{L}=667$~fb$^{-1}$ 
and $\mathcal{L}=416$~fb$^{-1}$, respectively.

For scenario B we now compare our results for $e^-e^-\to \tilde e^-_L \tilde e^-_R$ 
with the T-odd asymmetry $A_T$
studied in \cite{Bartl:2004jj} for neutralino production 
$e^+e^-\to \tilde\chi^0_1\tilde\chi^0_2$ followed by the three-body decay
$\tilde\chi^0_2\to \tilde\chi^0_1 \ell^+\ell^-$, $\ell=e,\mu$, at the ILC
operating at $\sqrt{s}=500$~GeV.
For the optimal choice of ($P_{e^-}$,$P_{e^+}$)=($-0.9$,$+0.6$) 
for the longitudinal beam 
polarizations we obtain
$A_T=0.0013$ and for the cross section of the combined
process $\sigma(e^+e^-\to \tilde\chi^0_1\tilde\chi^0_2)\cdot 
\sum _\ell B(\tilde\chi^0_2\to \tilde\chi^0_1 \ell^+\ell^-)=11.9$~fb.
The integrated luminosity necessary to measure 
this asymmetry in $e^+e^-\to \tilde\chi^0_1\tilde\chi^0_2$ at $3\sigma$ would
be $\mathcal{L}=4.5\times 10^5$~fb$^{-1}$.
This example illustrates the potential of the $e^-e^-$ 
mode for an identification
of CP violation in the neutralino sector.
In this context we remark that it may be the case 
that the reaction $e^+e^-\to \tilde\chi^0_1\tilde\chi^0_2$ is
kinematically not allowed because the threshold is too high, 
however, $e^-e^-\to \tilde e^-_L \tilde e^-_R$ is accessible.
In such a case the reaction $e^-e^-\to \tilde e^-_L \tilde e^-_R$ 
and the CP sensitive
observables defined in Eq.~\rf{eq:Obs} may be a suitable way to determine the
phases $\phi_{M_1}$.

\begin{figure}[t]
\setlength{\unitlength}{1mm}
\begin{center}
\begin{picture}(150,120)
 \put(-53,-110){\mbox{\epsfig{figure=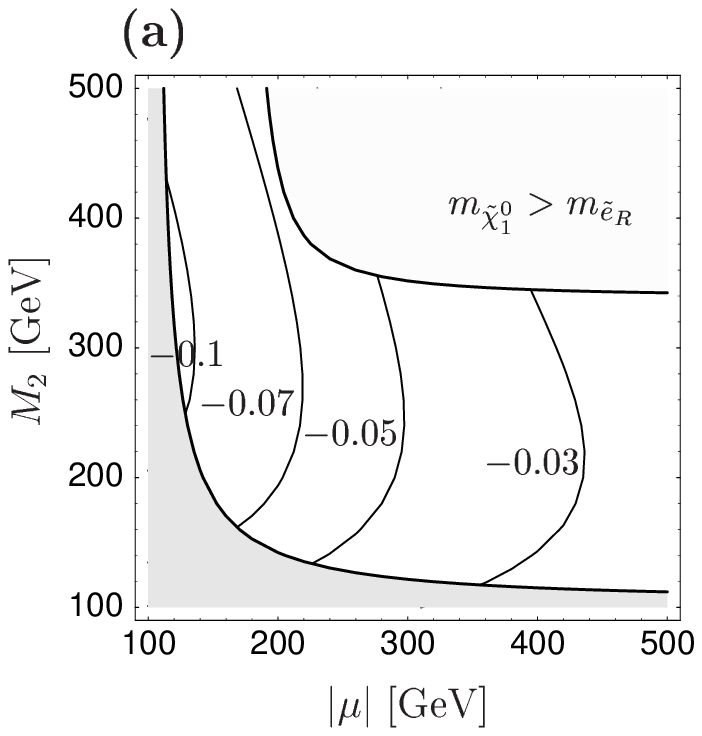,height=27cm,width=19.4cm}}}
 \put(27,-110){\mbox{\epsfig{figure=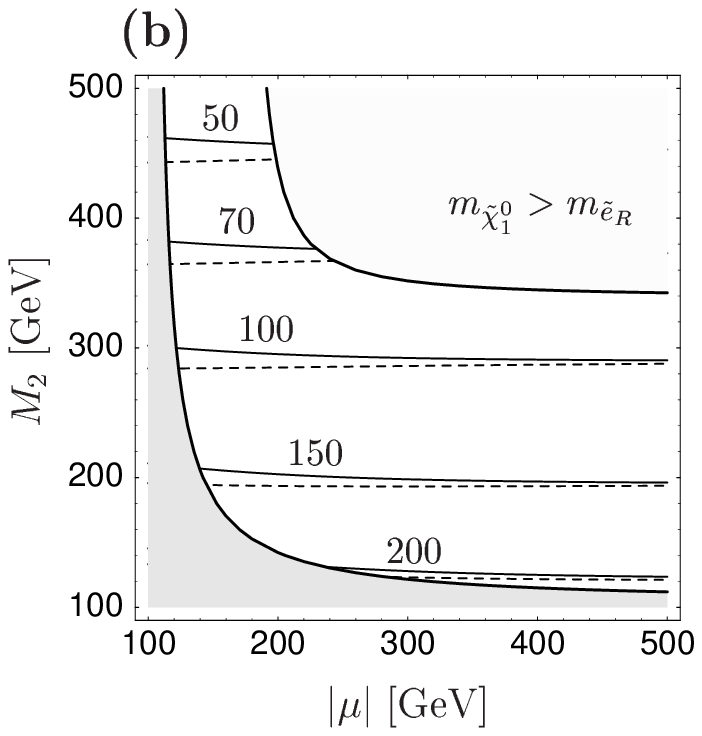,height=27cm,width=19.4cm}}}
 \put(25,128){\mbox{$\hat O[{\mathcal H}_2]~[{\rm fb}^{\frac{1}{2}}]$}}
 \put(100,128){\mbox{$\sigma(e^-e^-\to\ti e^-_L\ti e^-_R)~[{\rm fb}]$}}
\end{picture}
\end{center}
\vspace{-7cm}
\caption{Contour lines (a) of the effective CP observable 
$\hat O[{\mathcal H}_2]$ for $\phi_{M_1}=0.5\pi$  
and (b) of 
the corresponding cross section $\sigma(e^-e^-\to\ti e^-_L\ti e^-_R)$
for $\phi_{M_1}=0$ (dashed line) and $\phi_{M_1}=0.5\pi$ (solid line),
in the $|\mu|$--$M_2$ plane . The
parameters which are not varied are as given in scenario A 
of Table~\ref{tab1}. In the light-gray region $m_{\ti{\chi}^\pm_1}<104$~GeV 
and the region in the top right corner is excluded because
there $m_{\ti{\chi}^0_1}>m_{\ti{e}_R}$.
}
\label{fig:plot3}
\end{figure}
\begin{figure}[H]
\setlength{\unitlength}{1mm}
\begin{center}
\begin{picture}(150,120)
 \put(-53,-110){\mbox{\epsfig{figure=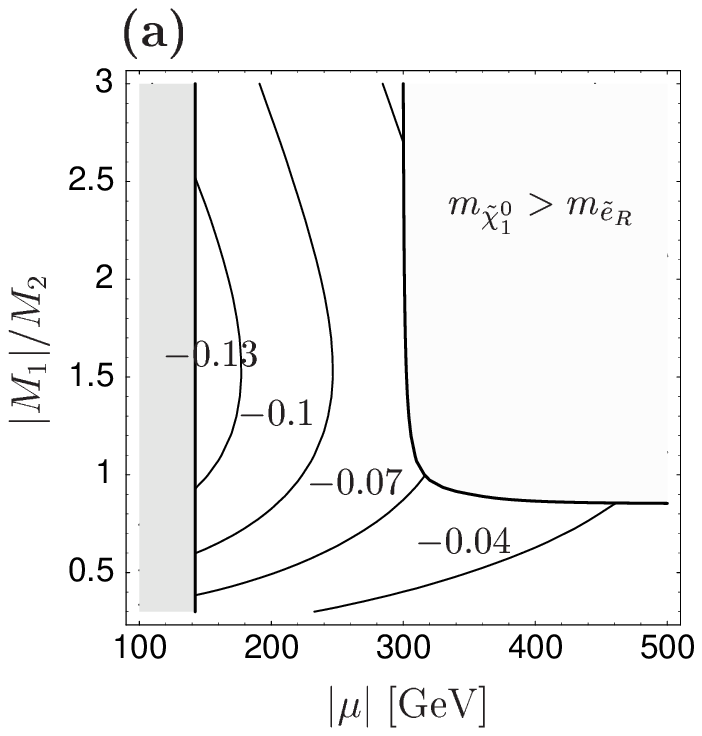,height=27cm,width=19.4cm}}}
 \put(27,-110){\mbox{\epsfig{figure=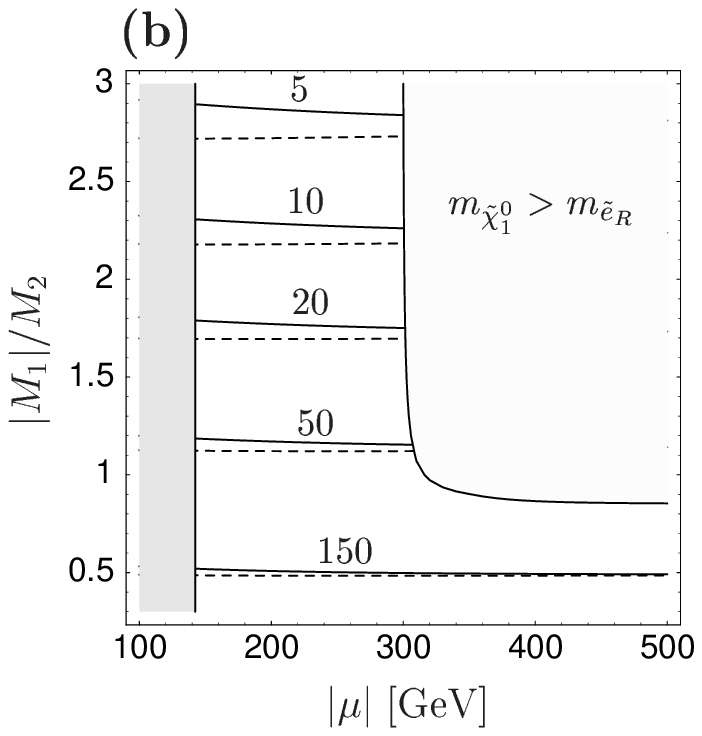,height=27cm,width=19.4cm}}}
 \put(25,128){\mbox{$\hat O[{\mathcal H}_2]~[{\rm fb}^{\frac{1}{2}}]$}}
 \put(100,128){\mbox{$\sigma(e^-e^-\to\ti e^-_L\ti e^-_R)~[{\rm fb}]$}}
\end{picture}
\end{center}
\vspace{-7cm}
\caption{Contour lines (a) of the effective CP observable 
$\hat O[{\mathcal H}_2]$ for $\phi_{M_1}=0.5\pi$
and (b) of 
the corresponding cross section $\sigma(e^-e^-\to\ti e^-_L\ti e^-_R)$
for $\phi_{M_1}=0$ (dashed line) and $\phi_{M_1}=0.5\pi$ (solid line)
in the $|\mu|$--$|M_1|/M_2$ plane with $M_2=200$~GeV. The
parameters which are not varied are as given in scenario A 
of Table~\ref{tab1}. In the light-gray region $m_{\ti{\chi}^\pm_1}<104$~GeV 
and the region in the top right corner is excluded because
there $m_{\ti{\chi}^0_1}>m_{\ti{e}_R}$.
}
\label{fig:plot4}
\end{figure}
%

\noindent {\bf c) Dependence on the gaugino/higgsino mass parameters}\\[.5em]
In Fig.~\ref{fig:plot3}a we show contour lines of the 
effective CP observable, Eq.~\rf{eq:EffAsy}, for
the weight function ${\mathcal H}_2$ in the 
$|\mu|-M_2$ plane, where the other parameters are as in scenario A
with $|M_1|=5/3 \tan^2 \Theta_W M_2$.
As one can see, the effective CP observable is larger
for $|\mu|\lsim M_2$, because the terms in the amplitude
squared which are not sensitive to CP violation are smaller
than the CP sensitive terms.
For the largest absolute value of the effective CP observable 
($|\hat O[{\mathcal H}_2]|=0.1$~fb$^{1/2}$) 
the integrated luminosity necessary to measure the corresponding CP 
sensitive observable $\langle {\mathcal H}_2\rangle$ at $3\sigma$
is $\mathcal{L}=888$~fb$^{-1}$.
Fig.~\ref{fig:plot3}b shows the associated production cross section
in the $|\mu|$--$M_2$ plane for scenario A and for comparison also for the
CP conserving case $\phi_{M_1}=0$.
As can be seen, the production cross section is almost independent of $|\mu|$,
because the leading contributions are due to the exchange of
neutralinos with dominant bino and wino components.
The production cross section, however, sensitively depends
on $|M_1|$, and decreases for increasing $|M_1|$, because then 
the heavier neutralino states are dominantly binos and winos.

In Fig.~\ref{fig:plot4}a we show the contour lines of the effective CP
observable $\hat O[{\mathcal H}_2]$ in the $|\mu|$--$|M_1|/M_2$ plane, fixing 
$M_2=200$~GeV. The remaining parameters are as in scenario A, 
see Table~\ref{tab1}.
The absolute value of the effective CP observable is increased
if the ratio $|M_1|/M_2$ is increased from a value of $0.5$~fb$^{1/2}$ 
to $1.5$~fb$^{1/2}$.
In order to probe $\langle {\mathcal H}_2\rangle$ at $3\sigma$,
an integrated luminosity of at least $\mathcal{L}=519$~fb$^{-1}$ 
($|\hat O[{\mathcal H}_2]|=0.13$~fb$^{1/2}$) is required in this case.
In Fig.~\ref{fig:plot4}b the production cross section for
the reaction $e^-e^-\to\ti e^-_L\ti e^-_R$ in the $|\mu|$--$|M_1|/M_2$ 
plane is displayed for $\phi_{M_1}=\frac{\pi}{2}$ and $\phi_{M_1}=0$.
Again the cross section is almost independent of the value of $|\mu|$
and decreases when $|M_1|/M_2$ is increased, since the leading contribution
is due to the exchange of $\tilde\chi^0_4$ which is mainly a bino.

\noindent {\bf d) Dependence on $\tan\beta$ and $\phi_\mu$}\\[.5em]
We have also studied the $\tan\beta$ and $\phi_\mu$ dependences of 
the CP sensitive observables. 
For larger values of $\tan\beta$ for the scenarios A and B
the effective CP observable is somewhat reduced, because in this case
the degree of the higgsino admixture to $\tilde\chi^0_1$ is decreased. 
The influence of $\phi_\mu$ on the CP sensitive observables is less 
strong, especially in scenario A. In order to understand this point 
qualitatively, one can use approximative fromulae of the neutralino 
mixing matrix elements (see e.g. \cite{Drees})
that enter the relevant coupling 
$\Im m\{f^{L*}_1 f^{L}_3 f^{R*}_1 f^{R}_3\}$, showing 
that the leading term is proportional to $\sin\phi_{M_1}$ 
and the $\sin\phi_\mu$ dependence is less pronounced.
Furthermore, the measurabilities of 
$\langle {\mathcal H}_1\rangle$ and 
$\langle {\mathcal H}_2\rangle$, Eq.~\rf{eq:Obs}, 
increase for smaller selectron masses 
in which case the partial cancellation of $t-$channel 
and $u-$channel contributions is smaller, see Eq.~\rf{eq:propadiff}.
 
\begin{figure}[t]
\setlength{\unitlength}{1mm}
\begin{center}
\begin{picture}(150,35)
\put(-53,-152.5){\mbox{\epsfig{figure=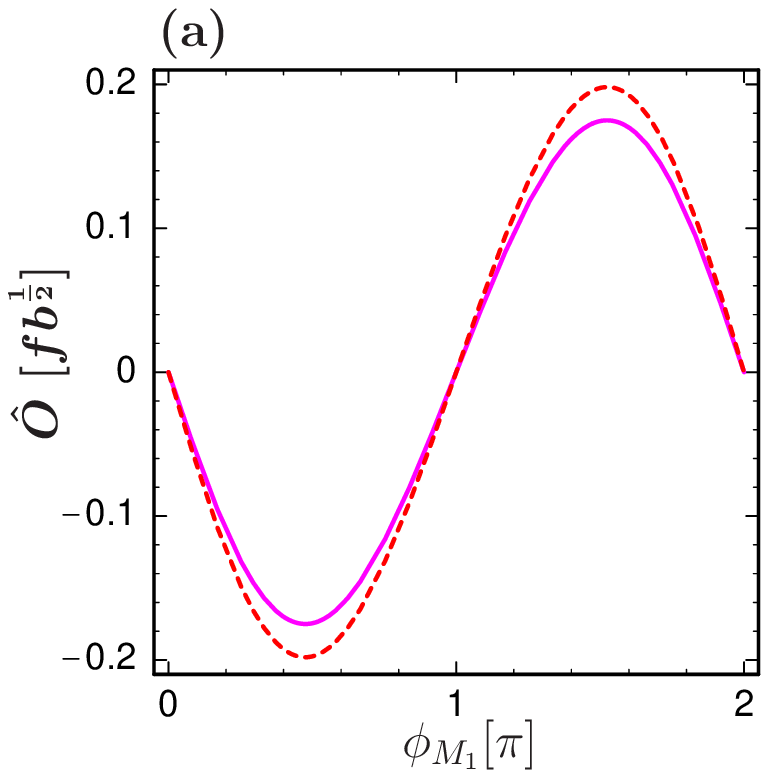,height=22.cm,width=19.4cm}}}
\put(30,-144){\mbox{\epsfig{figure=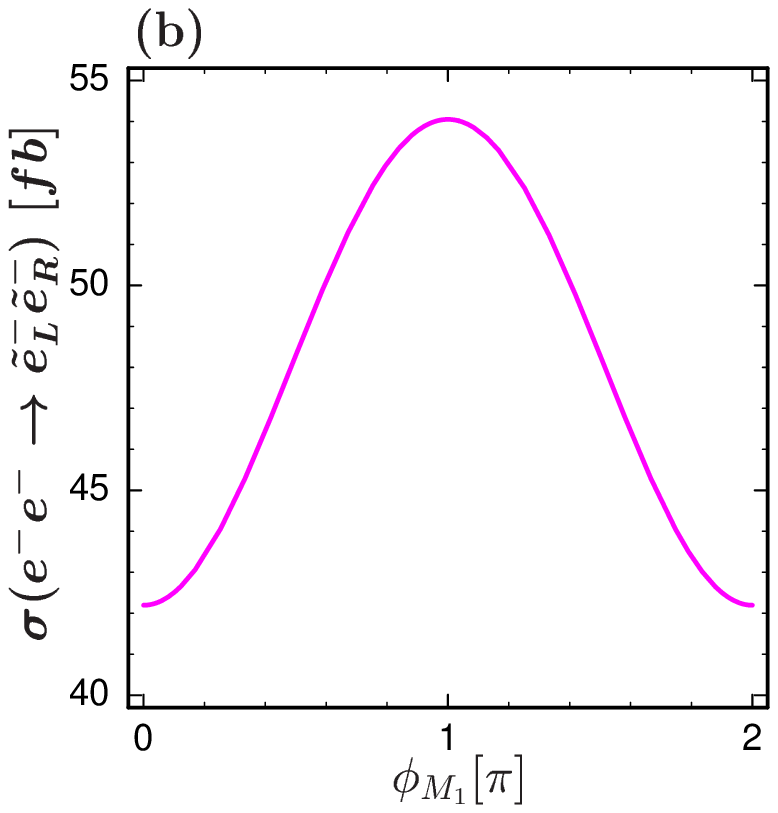,height=21cm,width=18.5cm}}}
\end{picture}
\end{center}
\caption{(a) Effective CP observable $\hat{O}[{\mathcal H}_i]$, Eq.~\rf{eq:EffAsy},
as a function of $\phi_{M_1}$ for scenario C 
of Table~\ref{tab1}, 
with ${\mathcal H}_1=\rm{sign}[\cos\theta \sin(\eta-2\phi)]$ (solid line) and
${\mathcal H}_2=\sin^2\theta \cos\theta\ \sin(\eta-2\phi)$ (dashed line)
and (b) the corresponding cross section 
$\sigma(e^-e^-\to\ti e^-_L\ti e^-_R)$. }
\label{fig:plot5}
\end{figure}

\noindent {\bf e) Scenario with light neutralinos $\tilde\chi^0_i$}\\[.5em] 
In Fig.~\ref{fig:plot5} we show the effective CP observables, 
Eq.~\rf{eq:EffAsy}, that are based on the weight functions ${\mathcal H}_1$ 
and ${\mathcal H}_2$ and the associated production cross section 
as a function of $\phi_{M_1}$ for scenario C, given in Table~\ref{tab1}.
In this scenario $\ti{\chi}^0_1$ and $\ti{\chi}^0_2$ are
mainly higgsinos, $\ti{\chi}^0_4$ is mainly a bino and $\ti{\chi}^0_3$ 
is mainly a wino with a pronounced bino admixture (see Table \ref{tab2}). 
Due to the moderate values of the heavier neutralino masses 
the $\ti{\chi}^0_3$--$\ti{\chi}^0_4$
interference term in Eq.~\rf{eq:Msqut} gives the leading contribution to
the CP sensitive observables $\langle {\mathcal H}_1\rangle$ and 
$\langle {\mathcal H}_2\rangle$. As can be seen in Fig.~\ref{fig:plot5}a, 
the effective CP observables reach their minimum (maximum) value
at $\phi_{M_1}\approx 0.5\pi~(1.5\pi)$. 
For these values the integrated luminosities necessary 
to probe the corresponding CP sensitive observables 
$\langle {\mathcal H}_1\rangle$ and 
$\langle {\mathcal H}_2\rangle$ at $3\sigma$ are
$\mathcal{L}=284$~fb$^{-1}$ and $\mathcal{L}=240$~fb$^{-1}$.

\begin{figure}[t]
\setlength{\unitlength}{1mm}
\begin{center}
\begin{picture}(150,45)
\put(-53,-152.5){\mbox{\epsfig{figure=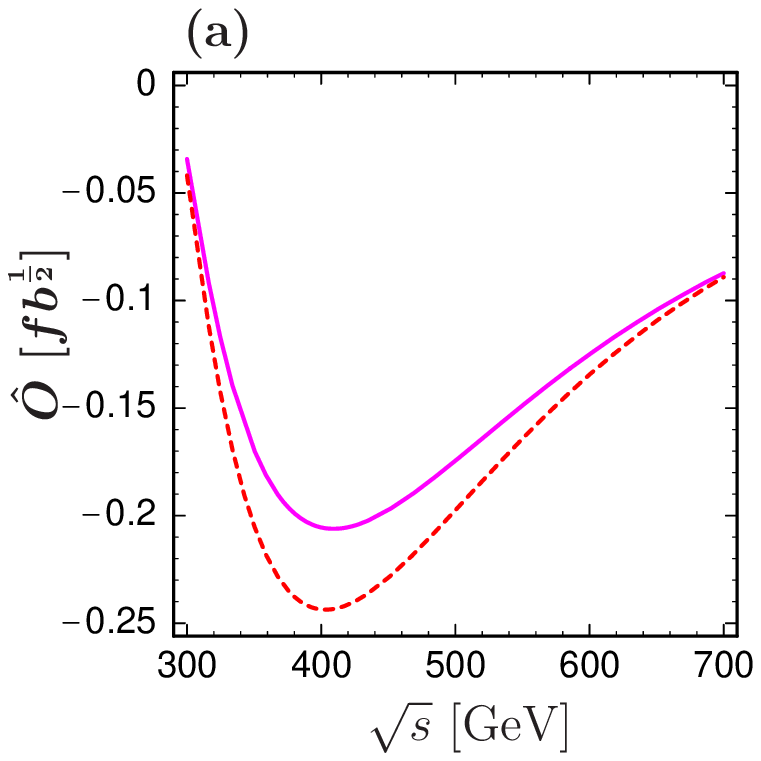,height=22.cm,width=19.4cm}}}
\put(32,-137){\mbox{\epsfig{figure=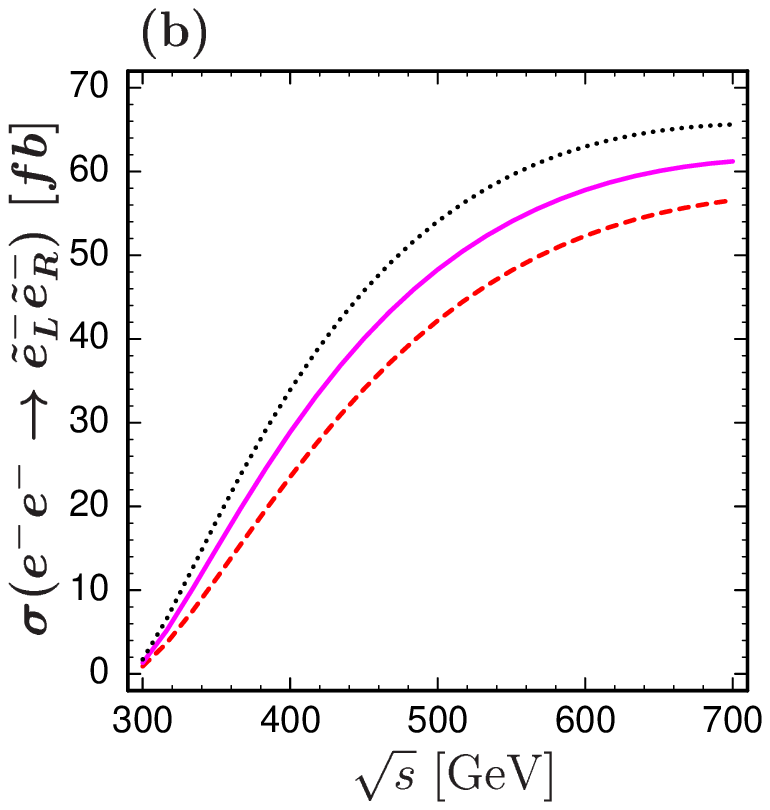,height=20.1cm,width=17.7cm}}}
\end{picture}
\end{center}
\caption{(a) Effective CP observable $\hat{O}[{\mathcal H}_i]$, Eq.~\rf{eq:EffAsy},
as a function of 
$\sqrt{s}$ for scenario C 
of Table~\ref{tab1}, 
with ${\mathcal H}_1=\rm{sign}[\cos\theta \sin(\eta-2\phi)]$ (solid line) and
${\mathcal H}_2=\sin^2\theta \cos\theta\ \sin(\eta-2\phi)$ (dashed line)
and (b) the corresponding cross section $\sigma(e^-e^-\to\ti e^-_L\ti e^-_R)$
for $\phi_{M_1}=0$ (dashed line), $\phi_{M_1}=0.5\pi$ (solid line) 
and $\phi_{M_1}=\pi$ (dotted line).}
\label{fig:plot6}
\end{figure}

In Fig.~\ref{fig:plot6}a the effective CP observables, Eq.~\rf{eq:EffAsy}, 
are plotted as a function of the 
center of mass energy $\sqrt{s}$ for scenario C.
As can be seen in Fig.~\ref{fig:plot6}a the minimum values of
the effective CP observables are reached at $\sqrt{s}\approx 400$~GeV.
At this point the integrated luminosities necessary to probe
$\langle {\mathcal H}_1\rangle$ and 
$\langle {\mathcal H}_2\rangle$ at $3\sigma$
decrease to $\mathcal{L}=206$~fb$^{-1}$ and $\mathcal{L}=156$~fb$^{-1}$
compared to the case $\sqrt{s}= 500$~GeV.
In Fig.~\ref{fig:plot6}b we show the 
$\sqrt{s}$ behavior of the production cross 
section $\sigma(e^-e^-\to\ti e^-_L\ti e^-_R)$ for scenario C 
including the CP conserving cases $\phi_{M_1}=0,\pi$ for comparison.

\section{Summary \label{conclusion}}

We have proposed and analyzed CP 
sensitive observables by means of the azimuthal angular
distribution of the produced selectrons at an $e^-e^-$ linear collider
with transverse beam polarizations. These observables are non-vanishing due to
the CP violating phases $\phi_{M_1}$ and $\phi_\mu$ in the neutralino sector.
We have numerically studied the MSSM parameter
dependence of these observables and of the production cross
section $\sigma(e^-e^- \to \ti{e}_L \ti{e}_R)$. Moreover,
we have also estimated the measurability of the proposed
CP sensitive observables. The best significances
(at the $3\sigma$ level) for their measurement are obtained
in scenarios where the GUT-inspired relation $|M_1|=5/3 \tan^2 \theta_W M_2$
does not hold. In such a case two exchanged neutralinos
can have a significant bino component, where the 
interference term of the corresponding amplitudes 
gives the dominant contribution to the CP sensitive observables.

\vskip10mm
\section*{Acknowledgements}

This work is supported by the 'Fonds zur F\"orderung der
wissenschaftlichen Forschung' (FWF) of Austria, project. No. P18959-N16.


\begin{thebibliography}{99}

\bibitem{Aguilar-Saavedra:2005pw}
  J.~A.~Aguilar-Saavedra {\it et al.},
  Eur.\ Phys.\ J.\ C {\bf 46} (2006) 43
  [arXiv:hep-ph/0511344].

\bibitem{Feng:1999zv}
  J.~L.~Feng,
  Int.\ J.\ Mod.\ Phys.\ A {\bf 15} (2000) 2355
  [arXiv:hep-ph/0002055];\\
%
  S.~Schreiber,
  Int.\ J.\ Mod.\ Phys.\ A {\bf 18} (2003) 2827;\\
%
  A.~De Roeck,
  arXiv:hep-ph/0311138;\\
%
  C.~A.~Heusch,
  Int.\ J.\ Mod.\ Phys.\ A {\bf 20} (2005) 7338.

\bibitem{TP}
R.~Budny,
Phys.\ Rev.\ D {\bf 14} (1976) 2969;\\
%
H.~A.~Olsen, P.~Osland and I.~Overbo,
Phys.\ Lett.\ B {\bf 97} (1980) 286;\\
%
K.~Hikasa,
Phys.\ Rev.\ D {\bf 33} (1986) 3203;\\
%
J.~L.~Hewett and T.~G.~Rizzo,
Z.\ Phys.\ C {\bf 34} (1987) 49 and  C {\bf 36} (1987) 209;\\
%
A.~Djouadi, F.~M.~Renard and C.~Verzegnassi,
Phys.\ Lett.\ B {\bf 241} (1990) 260;\\
%
C.~P.~Burgess and J.~A.~Robinson,
Int.\ J.\ Mod.\ Phys.\ A {\bf 6} (1991) 2707;\\
%
J.~Fleischer, K.~Kolodziej and F.~Jegerlehner,
Phys.\ Rev.\ D {\bf 49} (1994) 2174;\\
%
T.~G.~Rizzo,
JHEP {\bf 0302} (2003) 008
[arXiv:hep-ph/0211374]; JHEP {\bf 0308} (2003) 051
[arXiv:hep-ph/0306283];\\
%
M.~Diehl, O.~Nachtmann and F.~Nagel,
Eur.\ Phys.\ J.\ C {\bf 32} (2003) 17
[arXiv:hep-ph/0306247];\\
%
B.~Ananthanarayan and S.~D.~Rindani,
Phys.\ Rev.\ D {\bf 70} (2004) 036005
[arXiv:hep-ph/0309260]; Phys.\ Lett.\ B {\bf 606} (2005) 107
[arXiv:hep-ph/0410084]; JHEP {\bf 0510} (2005) 077
[arXiv:hep-ph/0507037]; arXiv:hep-ph/0601199;\\
%
B.~Ananthanarayan, S.~D.~Rindani, R.~K.~Singh and A.~Bartl,
Phys.\ Lett.\ B {\bf 593} (2004) 95
[arXiv:hep-ph/0404106];\\
%
S.~D.~Rindani,
Phys.\ Lett.\ B {\bf 602} (2004) 97
[arXiv:hep-ph/0408083].


\bibitem{TPsusy}
  S.~Y.~Choi, J.~Kalinowski, G.~Moortgat-Pick and P.~M.~Zerwas,
  Eur.\ Phys.\ J.\ C {\bf 22} (2001) 563
  [Addendum-ibid.\ C {\bf 23} (2002) 769]
  [arXiv:hep-ph/0108117];\\
%
  G.~Moortgat-Pick {\it et al.},
  arXiv:hep-ph/0507011.
%

\bibitem{TPchar}
  A.~Bartl, K.~Hohenwarter-Sodek, T.~Kernreiter and H.~Rud,
  Eur.\ Phys.\ J.\ C {\bf 36} (2004) 515
  [arXiv:hep-ph/0403265].

\bibitem{TPNeu1}
  A.~Bartl, H.~Fraas, S.~Hesselbach, K.~Hohenwarter-Sodek, T.~Kernreiter and G.~Moortgat-Pick,
  JHEP {\bf 0601} (2006) 170
  [arXiv:hep-ph/0510029].

\bibitem{TPNeu2}
  S.~Y.~Choi, M.~Drees and J.~Song,
  JHEP {\bf 0609} (2006) 064
  [arXiv:hep-ph/0602131].

\bibitem{Barger:2001nu}
V.~D.~Barger, T.~Falk, T.~Han, J.~Jiang, T.~Li and T.~Plehn,
Phys.\ Rev.\ D {\bf 64} (2001) 056007 [arXiv:hep-ph/0101106]
(and references therein).


\bibitem{Bartl:2003ju}
A.~Bartl, W.~Majerotto, W.~Porod and D.~Wyler,
Phys.\ Rev.\ D {\bf 68} (2003) 053005 [arXiv:hep-ph/0306050].

\bibitem{Bartl:2004jj}
  A.~Bartl, H.~Fraas, S.~Hesselbach, K.~Hohenwarter-Sodek and G.~A.~Moortgat-Pick,
  JHEP {\bf 0408} (2004) 038
  [arXiv:hep-ph/0406190].


\bibitem{Choi:1999cc}
  S.~Y.~Choi, H.~S.~Song and W.~Y.~Song,
  Phys.\ Rev.\ D {\bf 61}, 075004 (2000)
  [arXiv:hep-ph/9907474];\\
%
  A.~Bartl, H.~Fraas, O.~Kittel and W.~Majerotto,
  Phys.\ Rev.\ D {\bf 69}, 035007 (2004)
  [arXiv:hep-ph/0308141];
%
  Eur.\ Phys.\ J.\ C {\bf 36}, 233 (2004)
  [arXiv:hep-ph/0402016];
%
  Phys.\ Lett.\ B {\bf 598}, 76 (2004)
  [arXiv:hep-ph/0406309];
%
  Phys.\ Rev.\ D {\bf 70}, 115005 (2004)
  [arXiv:hep-ph/0410054];\\
%
  A.~Bartl, T.~Kernreiter and O.~Kittel,
  Phys.\ Lett.\ B {\bf 578}, 341 (2004)
  [arXiv:hep-ph/0309340];\\
%
  S.~Y.~Choi, M.~Drees, B.~Gaissmaier and J.~Song,
  Phys.\ Rev.\ D {\bf 69}, 035008 (2004)
  [arXiv:hep-ph/0310284];\\
%
  J.~A.~Aguilar-Saavedra,
  Nucl.\ Phys.\ B {\bf 697}, 207 (2004)
  [arXiv:hep-ph/0404104].

\bibitem{Drees}
  S.~Y.~Choi, M.~Drees and B.~Gaissmaier,
  Phys.\ Rev.\ D {\bf 70}, 014010 (2004)
  [arXiv:hep-ph/0403054].


\bibitem{Haber:1984rc}
H.~E.~Haber and G.~L.~Kane,
Phys.\ Rept.\  {\bf 117} (1985) 75.


\bibitem{Renard}
F.~M.~Renard, \it Basics of Electron Positron Collisions, 
\rm Editiors Frontieres, 
Dreux (1981).

\bibitem{Blochinger:2002zw}
  C.~Bl\"ochinger, H.~Fraas, G.~Moortgat-Pick and W.~Porod,
  Eur.\ Phys.\ J.\ C {\bf 24} (2002) 297
  [arXiv:hep-ph/0201282].

\bibitem{Thomas:1997ng}
  S.~D.~Thomas,
  Int.\ J.\ Mod.\ Phys.\ A {\bf 13} (1998) 2307
  [arXiv:hep-ph/9803420].


\bibitem{Atwood:1991ka}
  D.~Atwood and A.~Soni,
  Phys.\ Rev.\ D {\bf 45} (1992) 2405;\\
%
  M.~Diehl and O.~Nachtmann,
  Z.\ Phys.\ C {\bf 62} (1994) 397;\\
%
  B.~Grzadkowski and J.~F.~Gunion,
  Phys.\ Lett.\ B {\bf 350} (1995) 218
  [arXiv:hep-ph/9501339].


\end{thebibliography}
\end{document}